# Insight and Perspectives toward solving flaws of vegetation indices

Alin Khaliduzzaman[1,*], Satoshi Yamamoto[2], Yoh Nishimura[3]

[2]Specially Appointed Assistant Professor, Graduate School of Information Science, University of Hyogo, Hyogo 651-2197, Japan

[2]Associate Professor, Department of Bioresource Science, Akita Prefectural University, Akita 010-0444, Japan

[3]Professor, Agri-Innovation Education & Research Center, Akita Prefectural University, Akita 010-0451, Japan

*E-mail: khaliduzzaman@gsis.u-hyogo.ac.jp

**Abstract:** This perspective manuscript addressed several unsolved questions in vegetation index calculation such as the variation of the spectral fingerprint of crops and the differences in absorbance and reflectance spectral patterns of the young and mature leaves. The spectral shift is evident due to temporal and spatial variations. It means a generalized index, NDVI based on a near-infrared, and a red wavelength cannot precisely express the true meaning of a crops vegetation index. Thus commonly used vegetation indices have high possibility to undermine the actual photosynthetic capability or greenness of crops. Therefore, a crop specific vegetation index based on spectral characteristics in visible regions might be necessary to overcome this limitation of vegetation indices.

**One-Sentence Summary:** A single red and a near-infrared wavelength-based vegetation index undermines photosynthetic ability of the most of crops.

Is the spectral fingerprint the same for all crops and plants? Is the outlook color of crops the same? Are the absorbance and reflectance spectral patterns the same for young and mature leaves? If not, then why does the optical sensor measuring vegetation index (VI) use fixed and single wavelengths in the red region and why such VI will not undermine (underfit) the actual or true photosynthetic activity or greenness of the other crops?

Those above questions should become more compelling and catch the attention as more than 27 thousand Scopus indexed documents reported normalized difference vegetation index (NDVI), an average of 2000 papers each year over the last decade which comprises many scientific fields like earth and planetary sciences, environmental science, agricultural and biological sciences, computer science, engineering, social science, physics and astronomy, mathematics and material science, biochemistry, genetics, and molecular biology. Besides, more than 30 vegetation indices are reported in those literature to detect the presence and relative abundance of pigments in crops or plants as expressed in the solar-reflected optical spectrum. Among those indices, NDVI is the most popular and frequently used vegetation index in remote sensing, agriculture, and land-use studies. Therefore, this discussion proceeded with the reference of NDVI value and widely used DJI multispectral camera. This multispectral camera used a CMOS sensor sensitive to band pass wavebands with peaks at 650 nm and 840 nm for red and near-infrared light respectively. As the scientific basis of formulating NDVI is the relatively constant nature NIR reflectance and variable behavior of visible red light by the plant or plant parts during their growth stages, the greater value of index indicates the higher intensity of chloroplast or chlorophyll in parenchyma cells, thus an indicator of vegetation greenness (i.e., vegetation health and growth).

The visible color of crops varies due to the variation in reflectance behavior at visible range. Some crops (leaves) are dark greenish (e.g., onion leaves), and some are bright green (e.g., paddy leaves). If the peak of a reflectance spectra of the crop shifts towards blue bands; the crop looks dark green in the visible spectrum. On the other hand, crops look bright green when the peak shifts towards yellow bands in the visible spectrum. This shift can also occur as the absorbance bands vary with the leaf age and with the changes of form of combined chlorophyll *a* and *b* [1]. Hence, since the spectra are shifted with age of leaf (due to conversion of photo chlorophyll to chlorophyll), selection of wavelength for sensor design for vegetation index measurement may favor either young leaf or matured leaf. Therefore, spectral shift towards either shorter wavelength or longer wavelength must need to be considered in vegetation index calculation. It means a generalized index, NDVI based on a near-infrared (NIR), and a red (R) wavelength cannot precisely express the true meaning of a crops' vegetation index, thus underfit the actual greenness. Thus, there might be a high possibility to undermine the actual photosynthetic capability or greenness of crops. Therefore, a crop specific vegetation index based on spectral characteristics in visible regions might be necessary to overcome this limitation of common vegetation indices.

The plant leaf or chlorophylls absorb blue and red light and reflect green and NIR light [1]. We see leaf color as a dark green when it reflects light with more blue and less yellow or red parts. On the other hand, we see bright green leaves as it reflects light with more yellow or red parts. The difference in appearance (i.e., reflected light) between onion leaf (dark green) and paddy leaf (bright green) provides evidence of spectral shift which is clearly visible in fig. 1.

Although many researchers are thriving for a generalized vegetation index, no one addresses the concern of crop or plant specific spectral shift of reflected (or absorb band) light in VI calculation [2,3]. Milne and his co-researcher reported an interesting finding in their work "Unraveling the intrinsic color of chlorophyll" where they tried to prove that the spectral shift of photosynthetic pigments are higher than what we see [4]. It makes sense to support the necessity of reconsidering to design the formula of vegetation index.

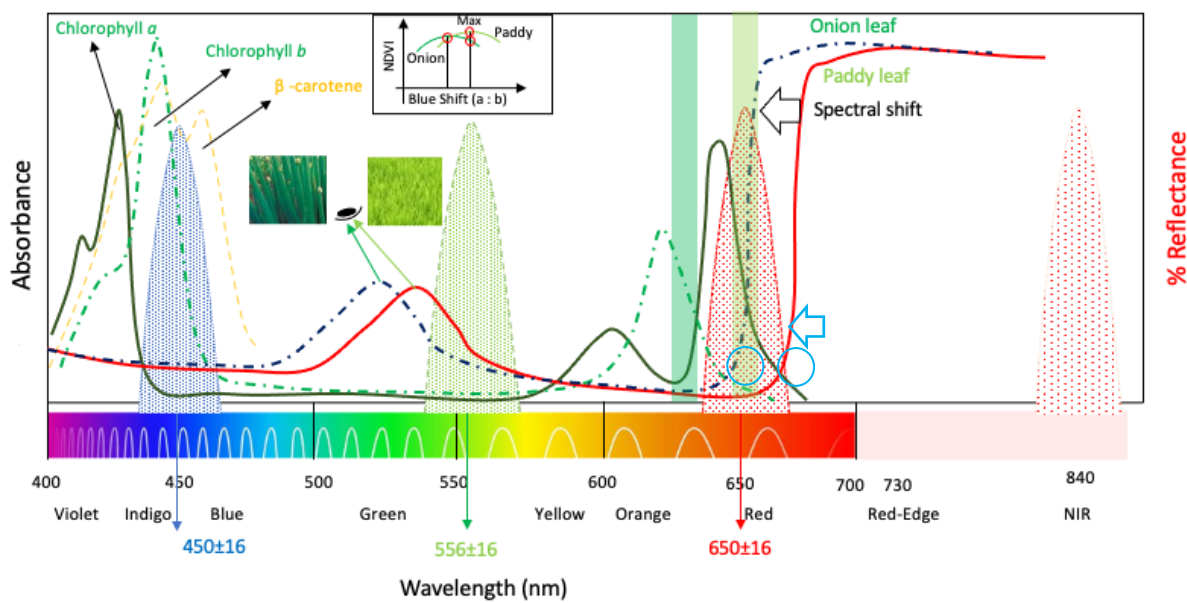

Figure 1: Schematic diagram explaining spectral variation of crops and camera sensitivity (CMOS sensor with optical filter) for vegetation index calculation.

The photosynthetic ability of a crop depends on the efficiency of light energy storage and utilization (i.e., how efficiently utilize the solar energy) by chlorophylls and ability to fix the $CO_2$ for photosynthetic carbon assimilation. This ability differs for C3 (e.g., Rice, Soyabean, wheat, oats, rye) and C4 plants (e.g., Corn, Onion, sorghum). Generally, the C4 plants have higher photosynthetic ability than C3 plants due to the presence of chloroplast in bundle sheath, hence the NDVI value of C4 crops should be higher at the same environmental conditions except for seasonal variation (e.g., winter). Therefore, a higher VI value of C3 plants than C4 plants in summer indicates the bias of wavelength selection in optical sensor design. As the nature, biochemical components and physiology differs crop to crop, their optical spectrum should be different in terms of peak position and intensity in visible and near infrared regions of electromagnetic waves. The concept of NDVI of vegetation (crops or plants) is based on spectral characteristics, maximum absorbance in visible red. The photosynthetic pigments Chlorophyll *a* and *b* absorb blue and red light (i.e., less reflectance) and reflect green and near-infrared light. But the dry plants reflect a greater part of visible red light. Thus, the NDVI value is mostly influenced by absorbance or reflectance behavior of visible red light. As the change in NIR light reflectance is lesser for growing plants, the spectral characteristics of red light of a plant or plant part mostly influence NDVI value. Thus, NDVI value is proportionally influenced by the red-light absorbance of crops. In that case, if visible red absorbance of crops increases, the reflectance will decrease and NDVI value will increase.

A new proposal of vegetation index or modified vegetation index named "Crop Specific Vegetation Index (CSVI)", *I* can be used to solve the flaws of NDVI without any hardware changes in devices used for NDVI calculation. The spectral shift can be solved in vegetation index calculation by addressing the concept of linear interpolation shown in eq.1.

$$I = \frac{NIR - (aG + bR)}{NIR + (aG + bR)}, a + b = 1 \ \& \ a, b \geq 0, \begin{cases} CSVI = NDVI, if\ a = 0 \\ spectral\ shift, if\ a > 0 \end{cases}$$

The coefficient *a* and *b* could be defined as spectral shift and intensity correction factors. The higher the value or weight *a*, the more linearly shifted toward the visible blue band. For more precise estimation, the spectral properties of leaf pigments such as chlorophyll a and b where the scientist should give more focus in the future. Including NDVI, there are more than 30 remote sensing indices found in literature. The above fact is also applicable to all those indices. Additionally, another concern should be taken into consideration to normalize the crop specific epicuticular waxy layer which reflects visible light especially in photosynthetic wavelength, 680 nm [5,6]. Such an inclusion in vegetation index calculation may contribute to better understanding of global change, biodiversity, and agriculture.